\documentclass[12pt]{article}
\begin{document}
\pagestyle{plain}
\setcounter{page}{1}
\begin{center}
{\large\bf The Cosmological Constant Problem and Nonlocal Quantum Gravity}
\vskip 0.3 true in {\large J. W. Moffat}
\vskip 0.3 true in {\it Department of Physics, University of Toronto,
Toronto, Ontario M5S 1A7, Canada}
\vskip 0.3 true in and
\vskip 0.3 true
in {\it Perimeter Institute for Theoretical Physics, Waterloo, Ontario N2J
2W9, Canada}

\date{\today}
\begin{abstract}%
A nonlocal quantum gravity theory is presented which is finite and unitary
to all orders of perturbation theory. Vertex form factors in Feynman
diagrams involving gravitons suppress graviton and matter vacuum
fluctuation loops by introducing a low-energy gravitational scale,
$\Lambda_{\rm Gvac} < 2.4\times 10^{-3}$ eV. Gravitons coupled to
non-vacuum matter loops and matter tree graphs are controlled by a vertex
form factor with the energy scale, $\Lambda_{GM}< 1-10$ TeV. A satellite
E\"otv\"os experiment is proposed to test a violation of the equivalence
principle for coupling of gravitons to pure vacuum energy compared to
matter. \end{abstract}

\vskip 0.3 true in
\end{center}
Talk given at the MRST 2002 Meeting, Perimeter Institute for Theoretical
Physics, May 1-17, 2002, Waterloo, Ontario, Canada, and at the XVIIIth IAP
Colloquium: Observational and Theoretical Results on the Accelerating
Universe, July 1-5, 2002, Paris, France.
\vskip 0.3 true in

\vskip 0.2 true in
Dedicated to the memory of George Leibbrandt.

\vskip 0.3 true in
e-mail: john.moffat@utoronto.ca

\section{\bf Introduction}

It is generally agreed that the cosmological constant problem is
one of the most severe problems facing modern particle and gravitational
physics. It is believed that its solution could significantly alter our
understanding of particle physics and cosmology~\cite{Straumann}.

There is now mounting observational evidence~\cite{Perlmutter} that
the universe is accelerating and that there exists some form of dark
energy. One possible explanation for the accelerating expansion of the
universe is a small cosmological constant corresponding to a vacuum energy
density, $\rho_{\rm vac} \sim (2.4\times 10^{-3}\,{\rm eV})^4$.

There have been many attempts to solve the cosmological constant
problem (CCP). Weinberg's theorem~\cite{Weinberg} disallows all adjustment
models involving extra fields such as a dynamical scalar field.
Higher-dimensional models of the brane-bulk type with finite volume
extra-dimensions do not avoid fine-tuning~\cite{Cline}.

Superstring theory (M-theory) has not yet provided a solution to the
CCP. This could be due to the problem of understanding how to introduce
supersymmetry breaking into string theory models, although there may be
some deeper reason for the failure.

In the following, I will describe a possible resolution of
the CCP, based on a model of a nonlocal
quantum gravity theory and field theory that suppresses the coupling of
gravity to vacuum energy density. The theory can be tested by performing
E\"otv\"os experiments on Casimir vacuum energy in satellites.

\section{\bf Gravitational Coupling to Vacuum Energy}

We can define an effective cosmological constant
\begin{equation}
\lambda_{\rm eff}=\lambda_0+\lambda_{\rm vac},
\end{equation}
where $\lambda_0$ is the ``bare'' cosmological
constant in Einstein's classical field equations,
and $\lambda_{\rm vac}$ is the contribution that arises from the
vacuum density $\lambda_{\rm vac}=8\pi G\rho_{\rm vac}$.

Already at the standard model electroweak scale $\sim 10^2$ GeV, a
calculation of the vacuum density $\rho_{\rm vac}$, based on local quantum field
theory, results in a discrepancy of order $10^{55}$ with the observational
bound
\begin{equation}
\label{vacbound}
\rho_{\rm vac} \leq 10^{-47}\, ({\rm GeV})^4.
\end{equation}
This results in a severe fine-tuning problem of order $10^{55}$,
since the virtual quantum fluctuations giving rise to $\lambda_{\rm vac}$
must cancel $\lambda_0$ to an unbelievable degree of accuracy.
This is the ``particle physics'' source of the cosmological
constant problem.

\section{\bf Nonlocal Quantum Gravity}

Let us consider a model of nonlocal gravity with the action
$S=S_g+S_M$, where ($\kappa^2=32\pi G$)
\footnote{The present version of a nonlocal quantum gravity and
field theory model differs in detail from
earlier published work~\cite{Moffat,Moffat2}. A paper is in preparation in which more complete
details of the model will be provided.}:
\begin{equation}
S_g=-\frac{2}{\kappa^2}\int d^4x\sqrt{-g}\biggl\{R[g,{\cal
G}^{-1}]+2\lambda_0\biggr\}
\end{equation}
and $S_M$ is the matter action, which for the simple case of
a scalar field $\phi$ is given by
\begin{equation}
S_M=\frac{1}{2}\int d^4x\sqrt{-g}\biggl(g^{\mu\nu}{\cal
G}^{-1}\nabla_\mu\phi{\cal F}^{-1}\nabla_\nu\phi
-m^2\phi{\cal F}^{-1}\phi\biggr).
\end{equation}
Here, ${\cal G}$ and ${\cal F}$ are nonlocal regularizing,
{\it entire} functions and $\nabla_\mu$ is the covariant
derivative with respect to the metric $g_{\mu\nu}$. As an example,
we can choose the covariant
functions
\begin{equation}
{\cal G}(x)=\exp\biggl[-{\cal D}(x)/\Lambda_G^2\biggr],
$$ $$
{\cal F}(x)=\exp\biggl[-({\cal D}(x)+m^2)/\Lambda_M^2)\biggr],
\end{equation}
where ${\cal D}\equiv\nabla_\mu\nabla^\mu$, and $\Lambda_G$ and
$\Lambda_M$ are gravitational and matter energy scales,
respectively~\cite{Moffat,Moffat2}.

We expand $g_{\mu\nu}$ about flat Minkowski spacetime:
$g_{\mu\nu}=\eta_{\mu\nu}+\kappa h_{\mu\nu}$. The propagators for the
graviton and the $\phi$ field in a fixed gauge are given by
\begin{equation}
{\bar D}^\phi(p)=\frac{{\cal G}(p){\cal F}(p)}{p^2-{\bar m}^2+i\epsilon},
\end{equation}
$$ $$
\begin{equation}
{\bar D}^G_{\mu\nu\rho\sigma}(p)=\frac{(\eta_{\mu\rho}\eta_{\nu\sigma}
+\eta_{\mu\sigma}\eta_{\nu\rho}-\eta_{\mu\nu}\eta_{\rho\sigma})
{\cal G}(p)}{p^2+i\epsilon},
\end{equation}
where ${\bar m}^2=m^2{\cal G}(p)$.

Unitarity is maintained for the S-matrix, because ${\cal G}$ and
${\cal F}$ are {\it entire} functions of $p^2$, preserving the Cutkosky
rules.

Gauge invariance can be maintained by satisfying certain
constraint equations for ${\cal G}$ and ${\cal F}$ in every order of
perturbation theory.  This guarantees that $\nabla_\nu T^{\mu\nu}=0$.

\section{\bf Resolution of the CCP}

In flat Minkowski spacetime, the sum of all {\it disconnected}
vacuum diagrams $C=\sum_nM^{(0)}_n$ is a constant factor in the
scattering S-matrix $S'=SC$. Since the S-matrix is unitary
$\vert S'\vert^2=1$, then we must conclude that $\vert
C\vert^2=1$, and all the disconnected vacuum graphs can be
ignored. This result is also known to follow from the Wick ordering of the field
operators.

Due to the equivalence principle {\it gravity couples to all
forms of energy}, including the vacuum energy density $\rho_{\rm vac}$, so we can no
longer ignore these virtual quantum fluctuations in the presence of a non-zero
gravitational field. Quantum corrections to $\lambda_0$ come from
loops formed from massive standard model (SM) states, coupled to external
graviton lines at essentially zero momentum.

Consider the dominant contributions to the vacuum
density arising from the graviton-standard model loop corrections.
We shall adopt a model consisting of a photon loop coupled to
gravitons, which will contribute to the vacuum polarization loop coorection
to the bare cosmological constant $\lambda_0$. The covariant photon action
is~\cite{Leibbrandt}:
\begin{equation}
S_A=-\frac{1}{4}\sqrt{-g}g^{\mu\nu}g^{\alpha\beta}{\cal
G}^{-1}F_{\mu\alpha}{\cal F}^{-1}F_{\nu\beta},
\end{equation} with
\begin{equation}
F_{\mu\alpha}=\partial_\mu A_\alpha-\partial_\alpha A_\mu.
\end{equation}

The lowest order correction to the graviton-photon vacuum loop will have
the form (in Euclidean momentum space):
\begin{equation}
\Pi^{\rm Gvac}_{\mu\nu\rho\sigma}(p)
=\kappa^2\int\frac{d^4q}{(2\pi)^4}V_{\mu\nu\lambda\alpha}(p,-q,-q-p)
$$ $$
\times{\cal
F}^\gamma(q^2)D^\gamma_{\lambda\beta}(q^2)V_{\rho\sigma\beta\gamma}(-p,q,p-q)
{\cal F}^\gamma((p-q)^2)D^\gamma_{\alpha\gamma}((p-q)^2){\cal G}^{\rm
Gvac}(q^2), \end{equation}
where $V_{\mu\nu\rho\sigma}$ is the
photon-photon-graviton vertex and in a fixed gauge:
\begin{equation}
D^\gamma_{\mu\nu}=-\frac{\delta_{\mu\nu}}{q^2}
\end{equation}
is the free photon propagator. Additional contributions to
$\Pi^{\rm Gvac}_{\mu\nu\rho\sigma}$ come from tadpole
graphs~\cite{Leibbrandt}.

This leads to the vacuum polarization tensor
\begin{equation}
\label{Ptensor}
\Pi^{\rm Gvac}_{\mu\nu\rho\sigma}(p)=\kappa^2
\int\frac{d^4q}{(2\pi)^4}\frac{1}{q^2[(q-p)^2]}
$$ $$
\times K_{\mu\nu\rho\sigma}(p,q)
\exp\biggl\{-q^2/\Lambda^2_M
-[(q-p)^2]/\Lambda^2_M-q^2/\Lambda^2_{{\rm Gvac}}\biggr\}.
\end{equation}
For $\Lambda_{{\rm Gvac}} \ll \Lambda_M$, we
observe that from power counting of the momenta in the loop integral, we
get
\begin{equation}
\Pi^{\rm {\rm Gvac}}_{\mu\nu\rho\sigma}(p)\sim
\kappa^2\Lambda_{{\rm Gvac}}^4N_{\mu\nu\rho\sigma}(p^2)
$$ $$
\sim\frac{\Lambda_{{\rm Gvac}}^4}{M^2_{\rm PL}}N_{\mu\nu\rho\sigma}(p^2),
\end{equation}
where $N(p^2)$ is a finite remaining part of $\Pi^{\rm {\rm Gvac}}(p)$ and
$M_{\rm PL}\sim 10^{19}$ GeV is the Planck mass.

We now have
\begin{equation}
\rho_{\rm vac}\sim M^2_{\rm PL}\Pi^{\rm {\rm Gvac}}(p)\sim\Lambda_{{\rm
Gvac}}^4.
\end{equation} If we choose $\Lambda_{{\rm Gvac}}\leq 10^{-3}$
eV, then the quantum correction to the bare cosmological constant
$\lambda_0$ is suppressed sufficiently to satisfy the observational bound
on $\lambda$, {\it and it is protected from large unstable radiative
corrections}.

This provides a solution to the
cosmological constant problem at the energy level of the standard model
and possible higher energy extensions of the standard model. The universal
fixed gravitational scale $\Lambda_{{\rm Gvac}}$ corresponds to the fundamental
length $\ell_{{\rm Gvac}}\leq 1$ mm at which virtual gravitational radiative
corrections to the vacuum energy are cut off.

The gravitational form factor ${\cal G}$, {\it when
coupled to non-vacuum SM gauge boson or matter loops}, will have the form
in Euclidean momentum space
\begin{equation} {\cal G}^{\rm GM}(q^2)
=\exp\biggl[-q^2/\Lambda_{GM}^2\biggr].
\end{equation}
If we choose $\Lambda_{GM} = \Lambda_{M}> 1-10$ TeV, then we will
reproduce the standard model experimental results, including the running
of the standard model coupling constants, and ${\cal G}^{GM}(q^2)={\cal
F}^M(q^2)$ becomes ${\cal G}^{GM}(0)={\cal F}^M(q^2=m^2)=1$ on the mass
shell. {\it This solution to the CCP leads to a violation of the WEP for
coupling of gravitons to vacuum energy and matter.} This
could be checked experimentally in a satellite E\"otv\"os experiment on the
Casimir vacuum energy~\cite{Ross}.

We observe that the required suppression of the vacuum diagram
loop contribution to the cosmological constant, associated with
the vacuum energy momentum tensor at lowest order,
demands a low gravitational energy scale $\Lambda_{{\rm Gvac}}\leq
10^{-3}$ eV, which controls the coupling of gravitons to
pure vacuum graviton and matter fluctuation loops.

In our finite, perturbative
quantum gravity theory nonlocal gravity produces a long-distance
infrared cut-off of the vacuum energy density through the low energy
scale $\Lambda_{{\rm Gvac}} < 10^{-3}$ eV~\cite{Moffat2}\footnote{The energy scale,
$\Lambda_G\sim 10^{-3}$ eV, has also
been considered by R. Sundrum and G. Dvali, G. Gabadadze and M.
Shifman~\cite{Sundrum}.}.
Gravitons coupled to {\it non-vacuum} matter tree graphs and matter loops
are controlled by the energy scale: $\Lambda_{GM}=\Lambda_{M} > 1-20$
TeV

The rule is: When external graviton lines are removed from a
matter loop, leaving behind {\it pure} matter fluctuation vacuum loops,
then those initial graviton-vacuum loops are suppressed by the form factor
${\cal G}^{{\rm Gvac}}(q^2)$ where $q$ is the internal matter loop momentum and
${\cal G}^{{\rm Gvac}}(q^2)$ is controlled by $\Lambda_{{\rm Gvac}} \leq 10^{-3}$ eV.
On the other hand, e.g. the proton first-order self-energy graph, coupled
to a graviton is controlled by $\Lambda_{GM}=\Lambda_M > 1-20$ TeV {\it
and does not lead to a measurable violation of the equivalence principle.}

The scales $\Lambda_M$ and $\Lambda_{{\rm Gvac}}$ are determined in
loop diagrams by the quantum non-localizable nature of the gravitons and
standard model particles. The gravitons coupled to matter and matter
loops have a nonlocal scale at $\Lambda_{GM}=\Lambda_M > 1-20$ TeV or a
length scale $\ell_M < 10^{-16}$ cm, whereas the gravitons coupled to
pure vacuum energy are localizable up to an energy scale
$\Lambda_{{\rm Gvac}}\sim 10^{-3}$ eV or down to a length scale $\ell_{{\rm Gvac}} > 1$
mm.

The fundamental energy scales $\Lambda_{{\rm Gvac}}$ and
$\Lambda_{GM}=\Lambda_M$ are determined by the underlying physical nature
of the particles and fields and do not correspond to arbitrary cut-offs,
which destroy the gauge invariance, Lorentz invariance and unitarity of the
quantum gravity theory for energies $>\Lambda_{{\rm Gvac}}\sim 10^{-3}$ eV.
The underlying explanation of these physical scales must be sought in a
more fundamental theory\footnote{It is interesting to note that if we
choose $\Lambda_{\rm GM}=\Lambda_M=5$ TeV, then we obtain
$\Lambda_{\rm Gvac}=\Lambda_M^2/M_{\rm PL}=2.1\times 10^{-3}$ eV.}

\section{\bf Conclusions}

We have described a possible solution to the cosmological
constant problem. The particle physics resolution requires that we
construct a nonlocal quantum gravity theory, which has vertex form factors
that are different for gravitons coupled to quantum {\it vacuum}
fluctuations and matter. This predicts a measurable violation of the WEP
for coupling to vacuum energy, but not to matter-graviton couplings or to
{\it non-vacuum matter loops}. This leads to a suppression of all standard
model vacuum loop contributions and, thereby, avoids a fine-tuning
cancellation between the ``bare'' cosmological constant $\lambda_0$ and
the vacuum contribution $\lambda_{\rm vac}$. It retains the experimental
agreement of the standard model and classical Einstein gravity. A
satellite E\"otv\"os experiment for Casimir vacuum energy could
experimentally decide whether nature does allow a vacuum energy WEP
violation.

Even though we can succeed in our nonlocal quantum gravity scenario to
explain why $\lambda_{\rm eff}$ is small, without excessive fine tuning, we
are still confronted with the ``coincidence'' problem associated with
dark energy and the existence of a small, positive cosmological
constant~\cite{Straumann}.

As a model of a future fundamental, nonlocal
quantum gravity theory, it does provide clues as to the resolution of the
``infamous'' cosmological constant problem.
\vskip 0.2 true in
{\bf Acknowledgments}
\vskip 0.2 true in
I thank Michael Clayton and George Gillies for helpful and
stimulating discussions. This work was supported by the Natural Sciences and
Engineering Research Council of Canada.  \vskip 0.2 true in

\end{document}